\documentclass[a4paper,11pt]{article}
\pdfoutput=1 

\usepackage{jinstpub} 

\title{Study of Streamer Development in Resistive Plate Chamber}

\author[a,b,1]{Jaydeep Datta \note{Corresponding author},}
\author[a,b]{Sridhar Tripathy,}
\author[a,b]{Nayana Majumdar}
\author[a,b]{and Supratik Mukhopadhyay}

\affiliation[a]{Saha Institute of Nuclear Physics, Sector 1, AF Block, Bidhan Nagar, Salt Lake, Kolkata 700064, India}
\affiliation[b]{Homi Bhabha National Institute, Training School Complex, Anushaktinagar, Mumbai 400094, India}

\emailAdd{jaydeep.datta@saha.ac.in}

\abstract{
This work has been carried out to simulate a Resistive Plate Chamber and corroborate it with experimental measurements in order to develop a numerical tool for studying the performance of the device for any gas mixture. This will allow us to explore the feasibility of operating these chambers in their avalanche mode within the Iron Calorimeter setup at India-based Neutrino Observatory with any eco-friendly substitute. The simulation has considered a hydrodynamic model of charge transport to emulate the electronic and ionic growths in the device as a function of the applied voltage which determines its working mode as either of the avalanche or streamer. In order to validate, the simulation result has been compared with compatible experimental data available in the literature.}


\arxivnumber{2005.13911}

\proceeding{XV Workshop on Resistive Plate Chambers and Related Detectors RPC2020\\
	10 - 14 February\\
	Rome}
\begin{document}
\maketitle
\flushbottom

\section{\label{sec:1}Introduction}
The magnetized Iron Calorimeter (ICAL) setup at India-based Neutrino Observatory (INO) \cite{INO2017} is designed to address the issue of neutrino mass hierarchy by studying the matter effect on neutrinos and anti-neutrinos traveling through the earth. It will have a modular structure consisting of 151 horizontal layers of iron plates of $50~\rm kt$ mass in total, interleaved with more than 29000 Resistive Plate Chambers (RPCs) leading to a setup of dimension $48 m \times 16 m \times 14.5 m$. The layers of RPC will be operated for tracking the muons produced by the charged current interaction of neutrino with the iron nuclei while passing through the ICAL for determination of their momentum, charge and flight path. The requirement of the experiment demands RPC operation in avalanche regime in order to achieve excellent position resolution and long-term operation. A gas mixture of R134a, iso-butane and SF$_6$ (95.5:4.2:0.3) has been chosen for this purpose.\\
All these gas components have high global warming potential (GWP) and hence the weighted GWP of the mixture crosses the permissible limit set by the Kyoto protocol \cite{Kyoto}. It calls for an exploration for an eco-friendly substitute, however, without compromising the performance of the RPC and the objectives of the experiment thereof. In this context, the authors have made an attempt to develop a simulation framework to study the performance of an RPC for a given gas mixture. This framework once validated would be useful for qualifying any proposed eco-friendly gas mixture and identifying the operating regime for avalanche mode performance of the RPC which is a prime requirement of the ICAL setup. Some preliminary results in this regard are available in \cite{Jaydeep}.\\
Many attempts have been made to simulate and understand the working principle of the avalanche and streamer modes of functioning of RPC. Most of them are based on either Monte Carlo methods \cite{ Riegler:2002vg, Lippmann2004, Salim2012, Bosnjakovic2014, Jash2016} or hydrodynamics \cite{Moshaii, Khorashad, Bosnjakovic}. A review of different simulation methods can be found in this reference \cite{Fonte2013}.\\
This report presents a simulation framework built on the basis of hydrodynamic model of electronic and ionic transports to calculate the streamer probability and efficiency of an RPC for a given gas mixture at different electric fields. It studies the growth of the charges in an RPC produced due to incident cosmic muons and their propagation at different applied voltages. The entire framework has been developed on the platform of a commercial Finite Element Method (FEM) package, COMSOL Multiphysics \cite{Comsol}. It has utilized relevant information of primary ionization (produced by HEED \cite{Heed}) and electron transport parameters in the gaseous medium (produced by MAGBOLTZ \cite{Magboltz}) that are required to carry out the simulation. To study the efficacy of the framework, the authors have considered the measured efficiency and streamer probability of an RPC operated with R134a and butane (97:3) gas mixture reported by \cite{Camarri1998} to compare with the simulated results.	 
 
\section{\label{sec:2} Model Geometry}
\begin{figure}[h]
	\centering
	\includegraphics[width=0.6\linewidth]{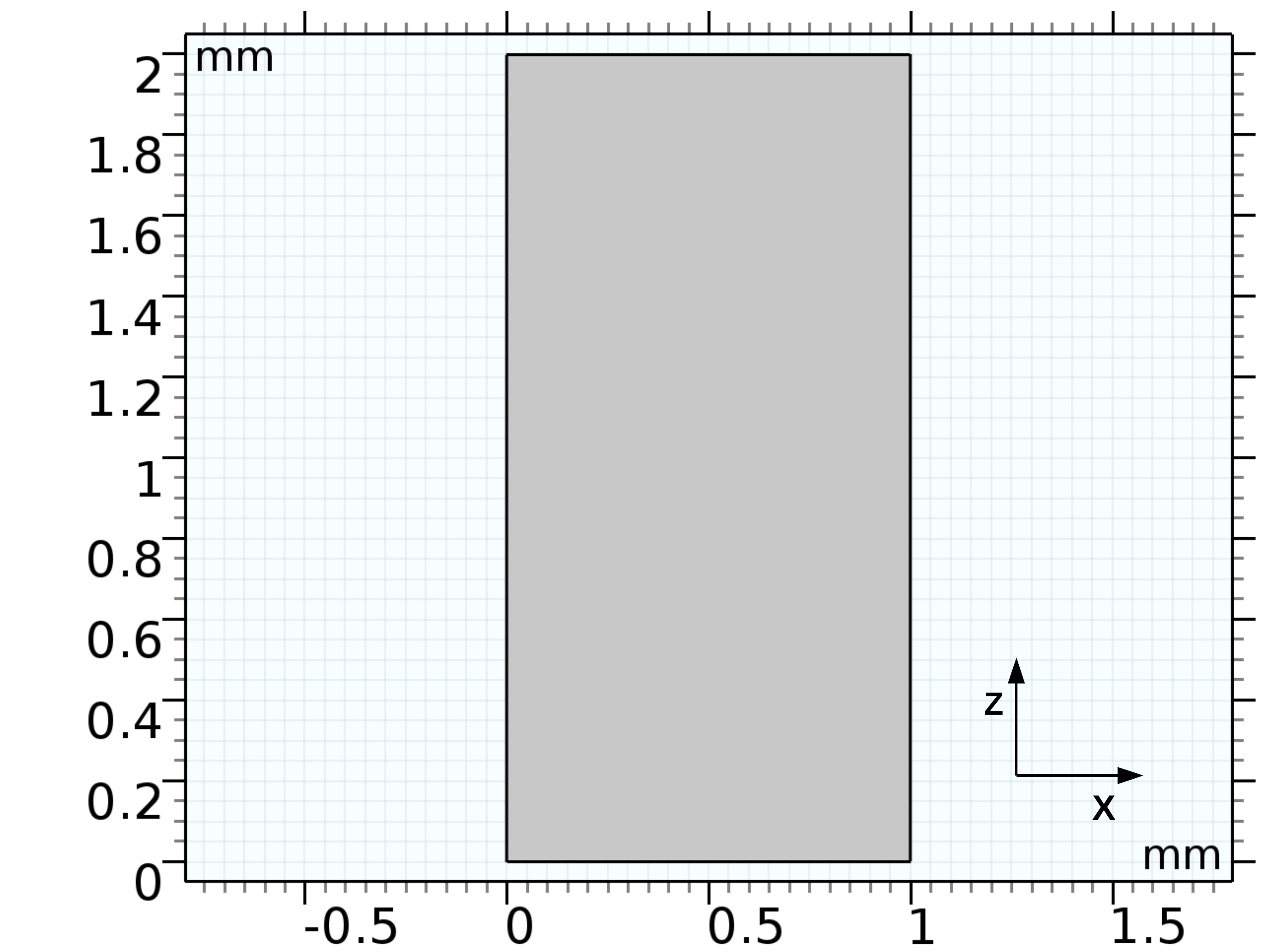}
	\caption{Geometry of the model}
	\label{fig:geom}
\end{figure}

3D modelling would obviously be the best choice for simulating the charge dynamics. However, it has not been adopted in the present work for its extensive computational expenses. Both of the 2D Cartesian and axisymmetric modelling approaches have their own limitations. In the Cartesian modelling, the charge growth is simulated in the 2D transverse cross-section (XZ-plane) of the detector active volume. On the other hand, the axisymmetric model simulates the growth on one side of the Z-axis and imposes rotational symmetry, hoping to achieve a pseudo-3D representation of the physical processes. \\
The RPC model has a gas gap (along Z-direction) of 2 mm, as illustrated in figure \ref{fig:geom}, filled with the gas mixture of R134a and butane (97:3) following the specifications mentioned in \cite{Camarri1998}. The electric field has been applied across the gas gap along the positive Z-direction. The length of the RPC has been considered 1 mm only along the X-direction. It is justified because the maximum radius of avalanche or streamer never exceeds $\frac{1}{\alpha}$ \cite{Raizer}, where $\alpha$ is the first Townsend coefficient. It is of the order of 10 mm$^{-1}$ for the gas mixture in consideration. The model has assumed geometrical symmetry along the Y-direction up to the length mentioned in the physics modules which is 1 mm in this case. 
\section{\label{sec:3} Simulation Method}
In the present work, the growth of the electronic charges in the gaseous medium through the multiplication of primary electrons has been simulated following the transport equations of hydrodynamics. It has been accomplished by calculating the charge at time steps optimized by COMSOL. This requires the electric field to be estimated at each time step which has been done by the "Electrostatic" module. It should be noted that in practical situation, the space charge can modify the electric field. The equations involved in calculating the electric field configuration $\vec{E}$, taking into account the space charge $\rho$ are given below,  
\begin{eqnarray}
\vec{E}=-\vec{\nabla} V\\
-\vec{\nabla}  d (\epsilon_0 \vec{\nabla} V - \vec{P}) = \rho
\end{eqnarray}
where V is the potential, $\vec{P}$ is the polarization vector, $\epsilon_0$ is the permittivity of the vacuum, and d is the thickness in Y-direction. The change of electric field has been considered in the X and Z-directions, while it has been considered constant in the Y-direction.\\
The growth of the electron and ion avalanches has been calculated using the "Transport of Dilute Species" module which assumes that the species (electron or ion) concentration is much smaller than that of the solvent (neutral gas molecules). The module has calculated transport of the species through diffusion and convection following the governing equations furnished below.
\begin{eqnarray}
\frac{\partial n_e}{\partial t} + \vec{\nabla} \cdot (-D \vec{\nabla} n_e + \vec {u}_e n_e) = S_e + S_{ph}\\
\frac{\partial n_i}{\partial t} + \vec{\nabla} \cdot (-D \vec{\nabla} n_i + \vec {u}_i n_i) = S_e + S_{ph}\\
S_e = (\alpha (\vec{E}) - \eta (\vec{E})) |\vec{u}_e| n_e (\vec{x}, t)\\
S_{ph} = Q_e \mu_{abs} \psi_0
\end{eqnarray}
where $n_e$ and $\ n_i$ denote electron and ion densities, respectively. The electron transport parameters in the gas medium have been expressed by $\vec{u}_e,\ D,\ \alpha$ and $\eta$ which represent the drift velocity, diffusion, the first Townsend and attachment coefficients, respectively. These are the functions of local electric field and have been calculated using MAGBOLTZ \cite{Magboltz}. Due to absence of significant magnetic field, the off-diagonal terms of diffusion tensor have become zero. This has led to a $2\ \times\ 2$ diagonal matrix for diffusion where the diagonal terms are the diffusion coefficients along the X and Z-directions. The terms $S_e$ and $S_{ph}$ represent the numbers of electrons produced due to gaseous ionization and photo-ionization, respectively. While $S_e$ depends upon the transport parameters and the electron density, $S_{ph}$ is dependent upon $Q_e$, $\mu_{abs}$, and $\psi_0$ where $Q_e$ is the quantum efficiency of the gas for electron generation from photo-ionization, $\mu_{abs}$ is the photon absorption coefficient of the same and $\psi_0$ is the photon flux.\\
The absorption of electrons in the anode has been taken care of by assuming drift of the electrons through the anode and similar condition has been set about the ions to take care of their absorption at the cathode. To incorporate the phenomenon of electron and ion diffusing and drifting out of the simulated volume, the two boundaries other than the cathode and the anode have been assumed open for them.\\
Photo-ionization of gas molecules, that takes place simultaneously with the Townsend ionization, plays an important role in the occurrence of the streamer \cite{Raizer}. To account for the photo-ionization contribution in the simulation, the cross-section of the same for different gas components has been taken into account from relevant sources \cite{Orlando1991, Lombos1967}. The work done by Capeill{\`{e}}re et.al. \cite{Capeill} shows that photon propagation in the gas volume can be described by the following equation. 
\begin{equation}\label{photon_propagation}                   
\vec{\nabla} (-c\vec{\nabla} \psi_0) + a \psi_0 = f\\            
\end{equation} where
\begin{eqnarray*}
	c = \frac{1}{3} \mu_{abs}\,,
	\qquad
	f = \delta S_e\,,
	\qquad
	a = \mu_{abs}
\end{eqnarray*}
Here, $\delta$ is the number of excited neutral molecules for each ionized molecule. In the present work, the photon propagation has been accounted for using "Coefficient Form Partial Differential Equation" module, which has been used to solve the equation \ref{photon_propagation}. As the electrodes are made of material which does not have scintillating property, the photon flux at the electrodes will be zero. This has been used as Dirichlet condition in the model. Like the charges, the photon propagation out of the simulation volume has been taken into account by considering the two boundaries other than the electrodes open for them.\\

\subsection{\label{sec:3a} Event Generation}
\begin{figure}[h]
	\centering
	\begin{minipage}{0.5\linewidth}
		\centering
		\includegraphics[width=\linewidth]{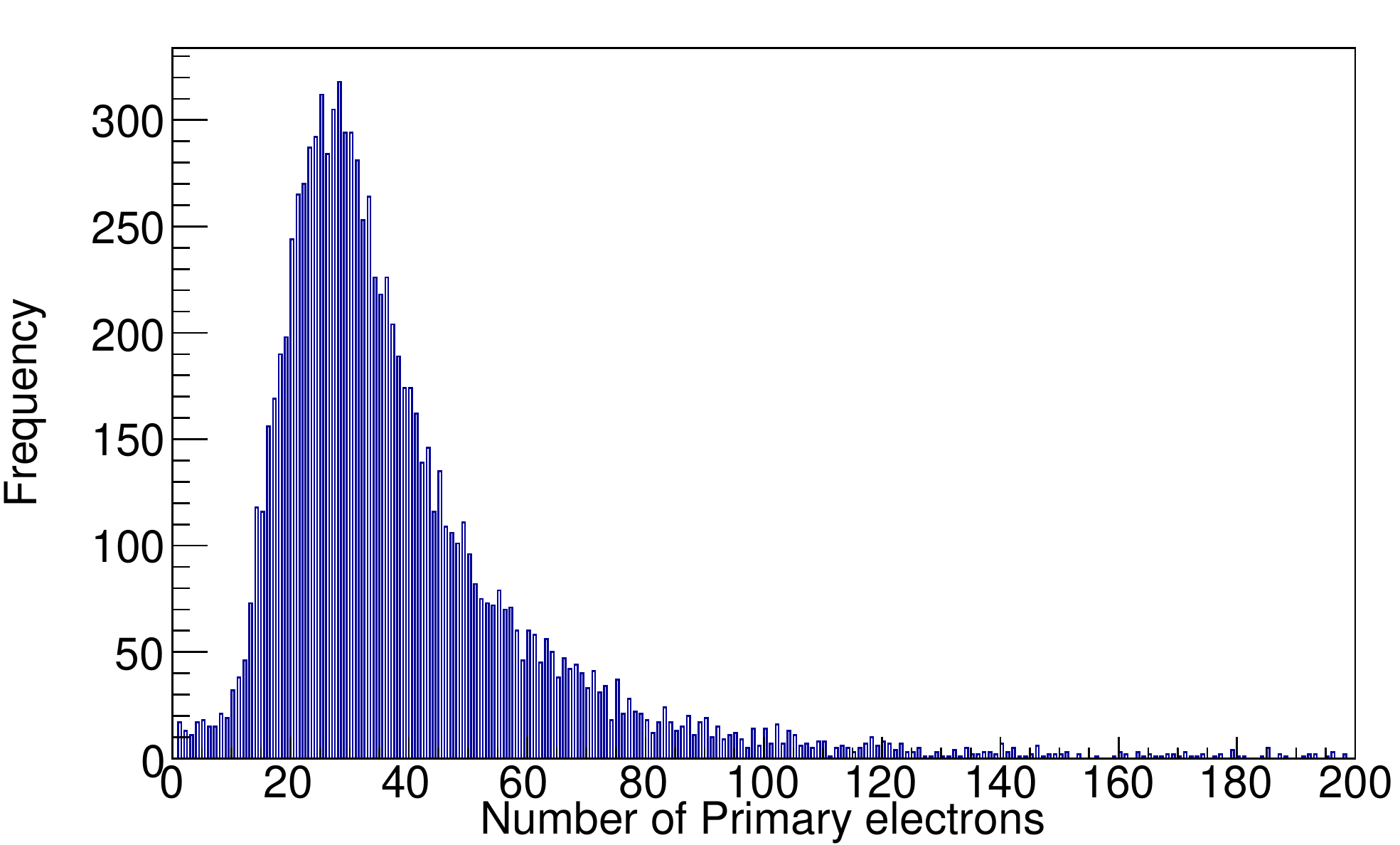}
		\caption{Number of primary electrons obtained from HEED}
		\label{fig:ne}
	\end{minipage}
	\begin{minipage}{0.4\linewidth}
		\centering
		\includegraphics[width=\linewidth]{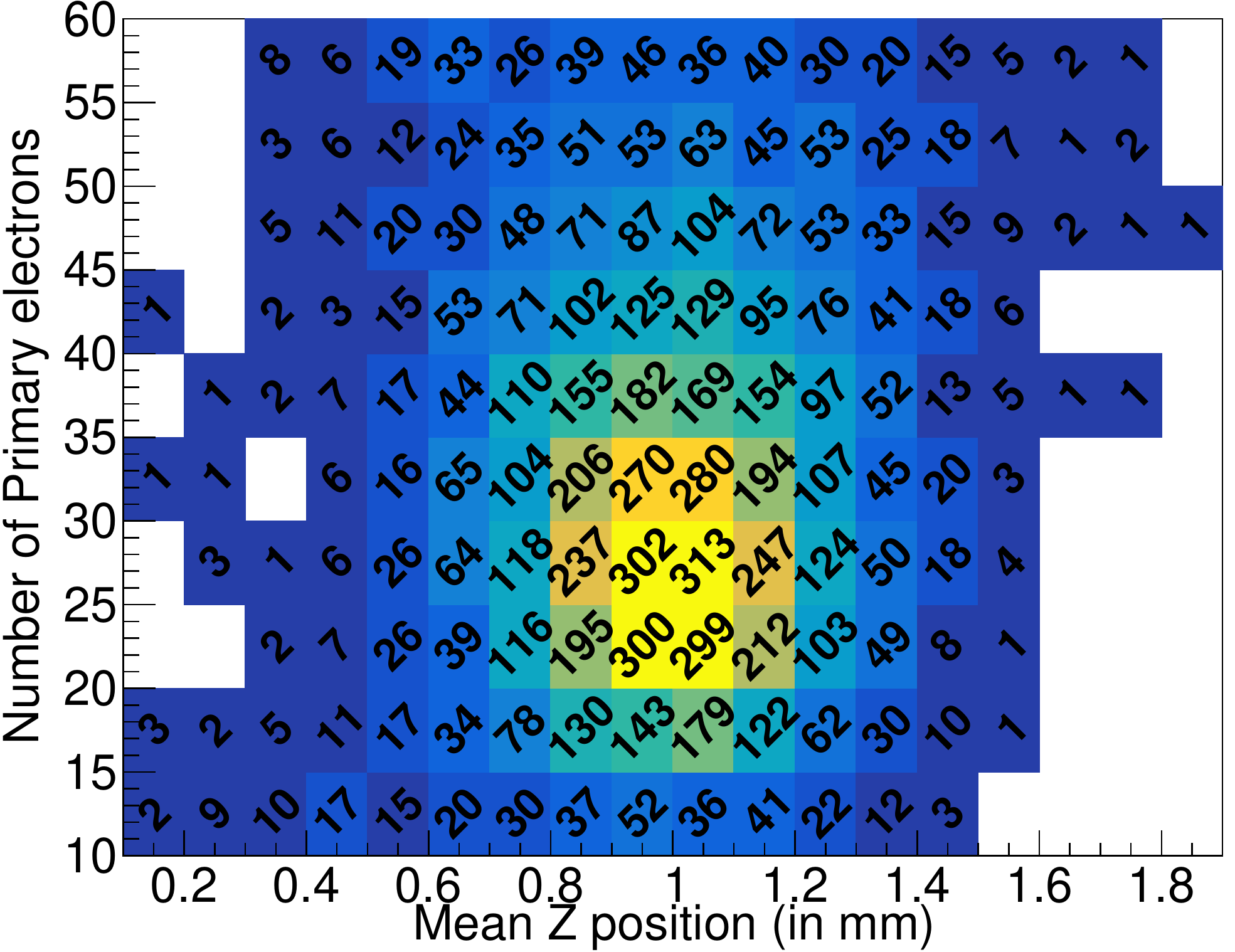}
		\caption{Variation of primary electron number with mean Z-position}
		\label{fig:epos}
	\end{minipage}	
\end{figure}

A total number of 10000 cosmic muons has been used for simulating the RPC response for each case of different applied voltages. Their energy has been varied from 1 to 10 GeV and the flux has been calculated following modification of Gaisser parameterization \cite{Gaisser} by Tang et.al. \cite{Tang}. The directionality of the muons has followed the zenith angle coverage (around 0$^\circ$) of the experimental setup as mentioned in \cite{Camarri1998}.\\
The primary ionization in the gas gap of the RPC caused by these muons has been simulated using HEED \cite{Heed}. It has provided information on the number of primary electrons and the position of the clusters formed by them for each muon event. The histogram of the primary electrons for these 10000 events has been shown in figure \ref{fig:ne} as obtained from HEED. These pieces of information have been utilized to recreate seed charge clusters for simulating their growth with the present hydrodynamic model.\\
It is evident that the events for which the average cluster size, total number of clusters and their range of spatial distribution along the Z-direction are same, will have same number of primary electrons. The weighted mean Z-position of all the clusters produced in each event will also be nearly same with very little variation. These events with same number of primary electrons and similar mean Z-positions have been considered as a group which has been represented by a seed charge cluster as mentioned earlier. Instead of simulating the growth of all the clusters generated in each event, simulation of the representative seed charge cluster has been carried out for each of the group of events. This has reduced the computation expense significantly. All the cosmic muon events thus have been classified according to their mean Z-position in a bin of 0.1 mm over a range of 0.1 - 1.9 mm and the number of primary electrons in a bin of 5 over a range of 10 - 60. It can be seen from figure \ref{fig:ne} that the said range of primary electron numbers has covered more than 90$\%$ events except some smaller and bigger clusters which are very few in frequency. This has led to reduced computation (by 50$\%$) at the cost of 2-3$\%$ error in estimating avalanche and streamer events. A 2D histogram of the seed clusters produced by grouping all the cosmic muon events (about 9000) is shown in figure \ref{fig:epos} in terms of their mean Z-position and the primary electrons. The numbers in figure \ref{fig:epos} denotes the frequency of events falling in each group.\\
The seed cluster has been represented by a Gaussian distribution \cite{Moshaii} with its mean position same as the mean Z-position calculated for the group of events. As the simulation has considered a 2D model in the XZ-plane, a two-variable Gaussian distribution has been used. The standard deviation for Z-variable of the Gaussian has been kept fixed for each mean Z-position with the condition that the 5$\sigma$ of the distribution always remains bound by the electrodes. It has determined the number of primary electrons in the seed cluster. As the experimental setup mentioned in \cite{Camarri1998} mostly detects vertical muons, the mean of the X-variable has been considered to be 0.5 mm, which is at the middle of the model geometry. For events with given mean Z-position and number of primary electrons, the hydrodynamic model deterministically leads to either avalanche or streamer discharge.\\

\subsection{\label{sec:3b}Stopping Conditions}
\begin{figure}[h]
	\centering
	\begin{minipage}{0.5\linewidth}
		\centering
		\includegraphics[width=\linewidth]{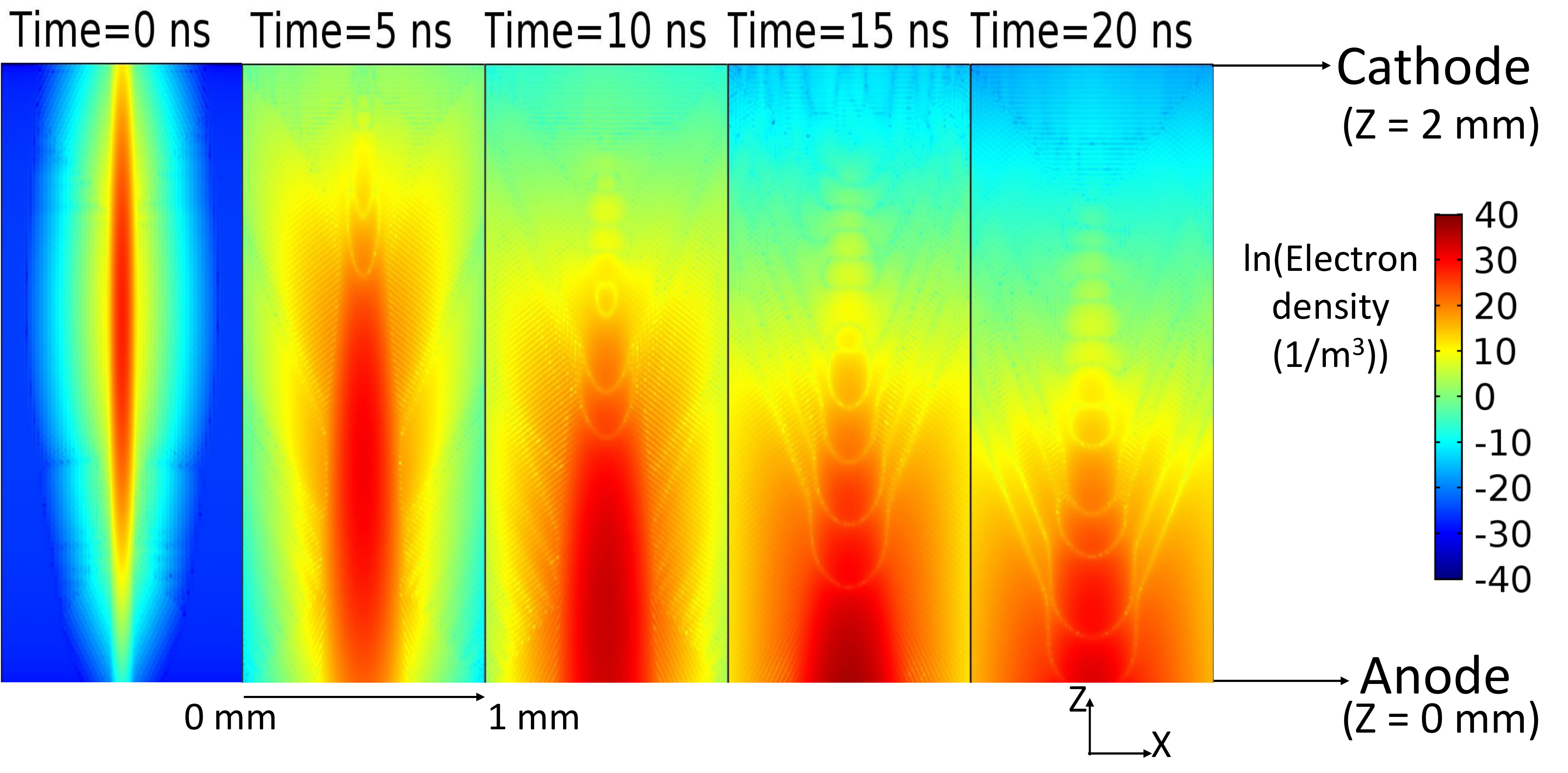}
		\caption{Growth of electronic charges in RPC for avalanche at 41 kV/cm for 10 primary electrons with mean Z-position of 1.2 mm}
		\label{fig:aval_grow}
	\end{minipage}
	\begin{minipage}{0.49\linewidth}
		\centering
		\includegraphics[width=\linewidth]{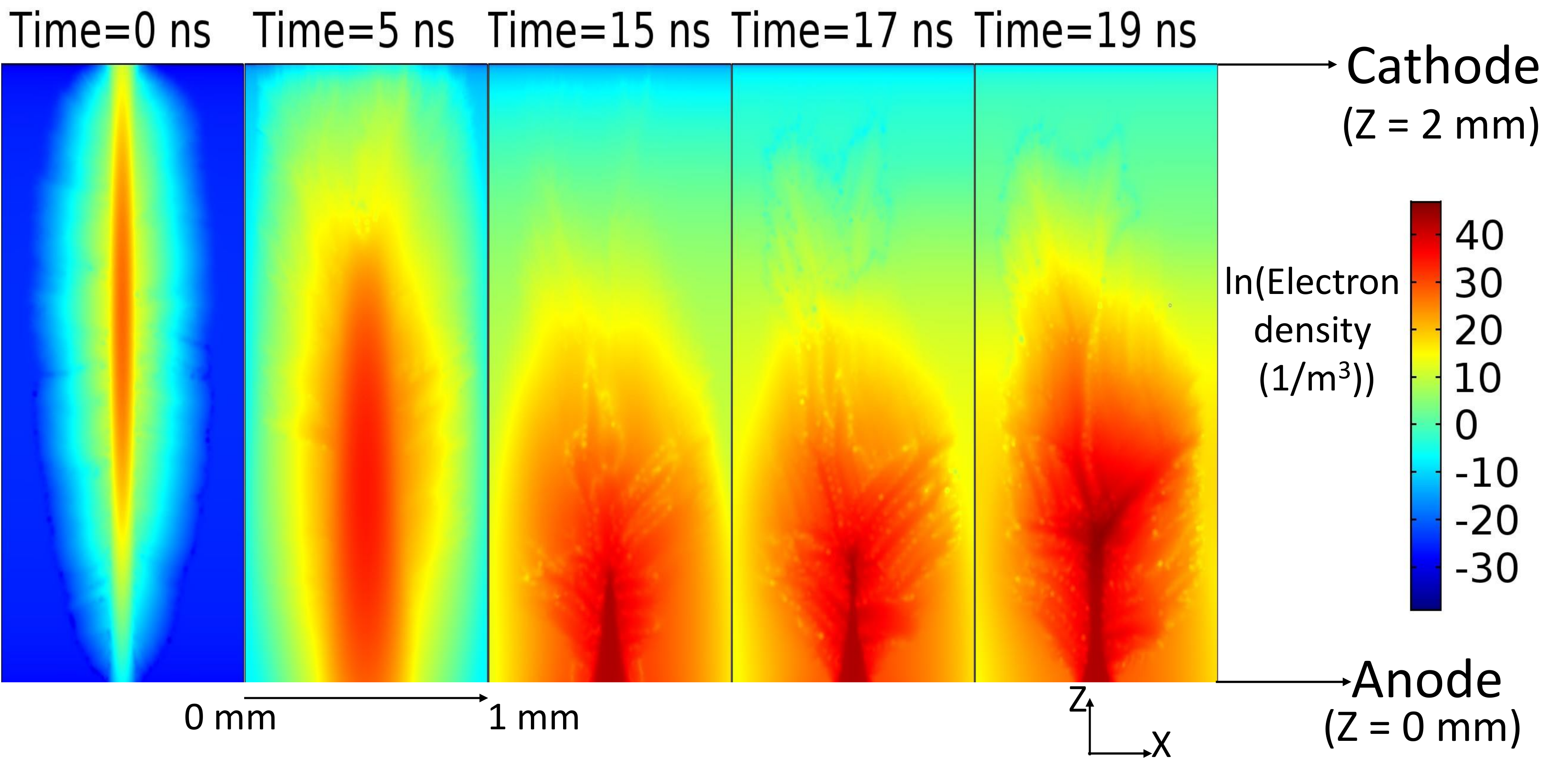}
		\caption{Growth of electronic charges in RPC for streamer at 47 kV/cm for 10 electrons with mean Z-position of 1.2 mm}
		\label{fig:str_grow}
	\end{minipage}   
\end{figure}

\begin{figure}[h]
	\centering
	\includegraphics[width=\linewidth]{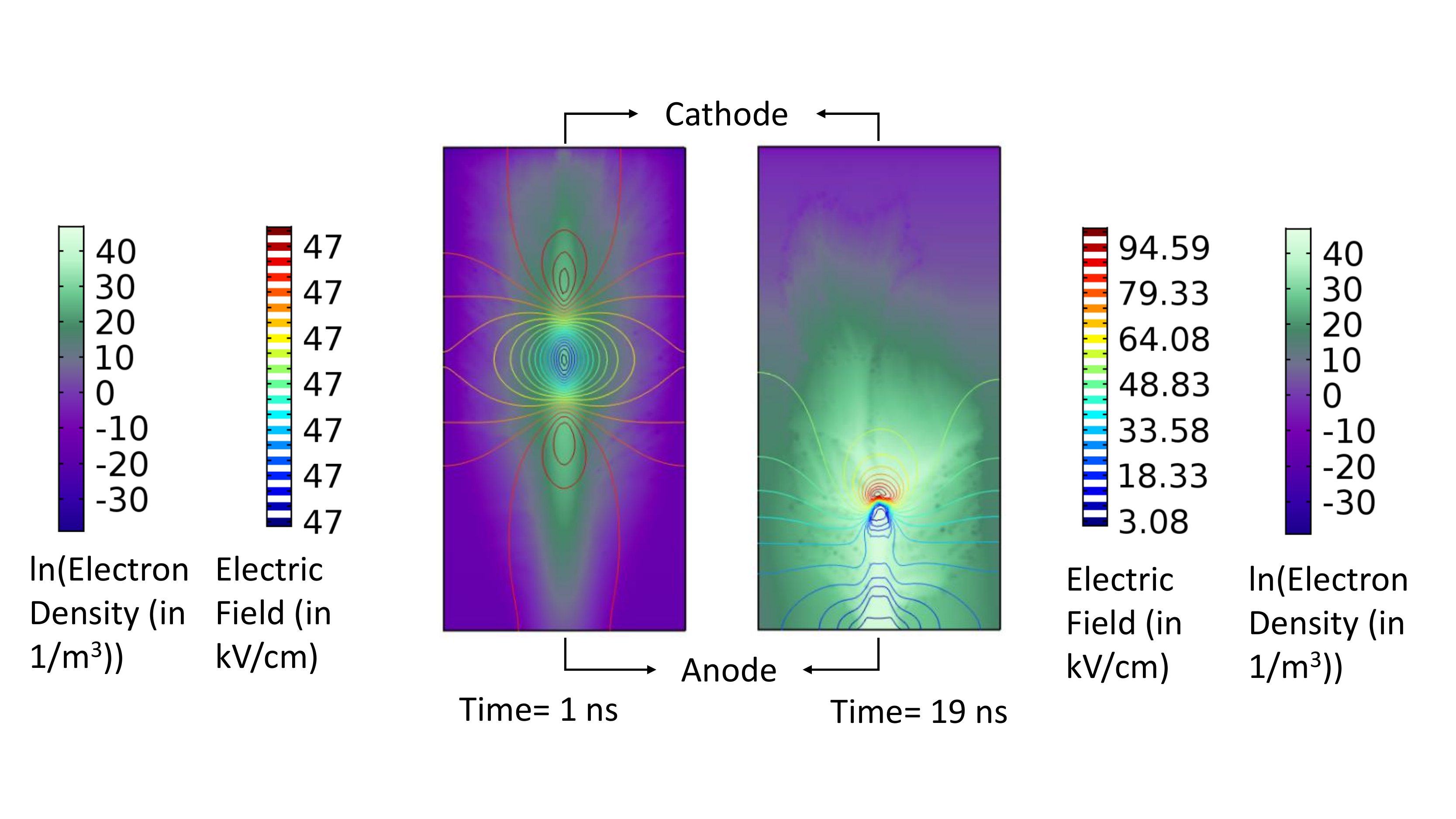}
	\caption{Time evolution of electric field and electron density at applied field 47 kV/cm}
	\label{fig:efld}
\end{figure}

\begin{figure}[h]
	\centering
	\includegraphics[width=0.6\linewidth]{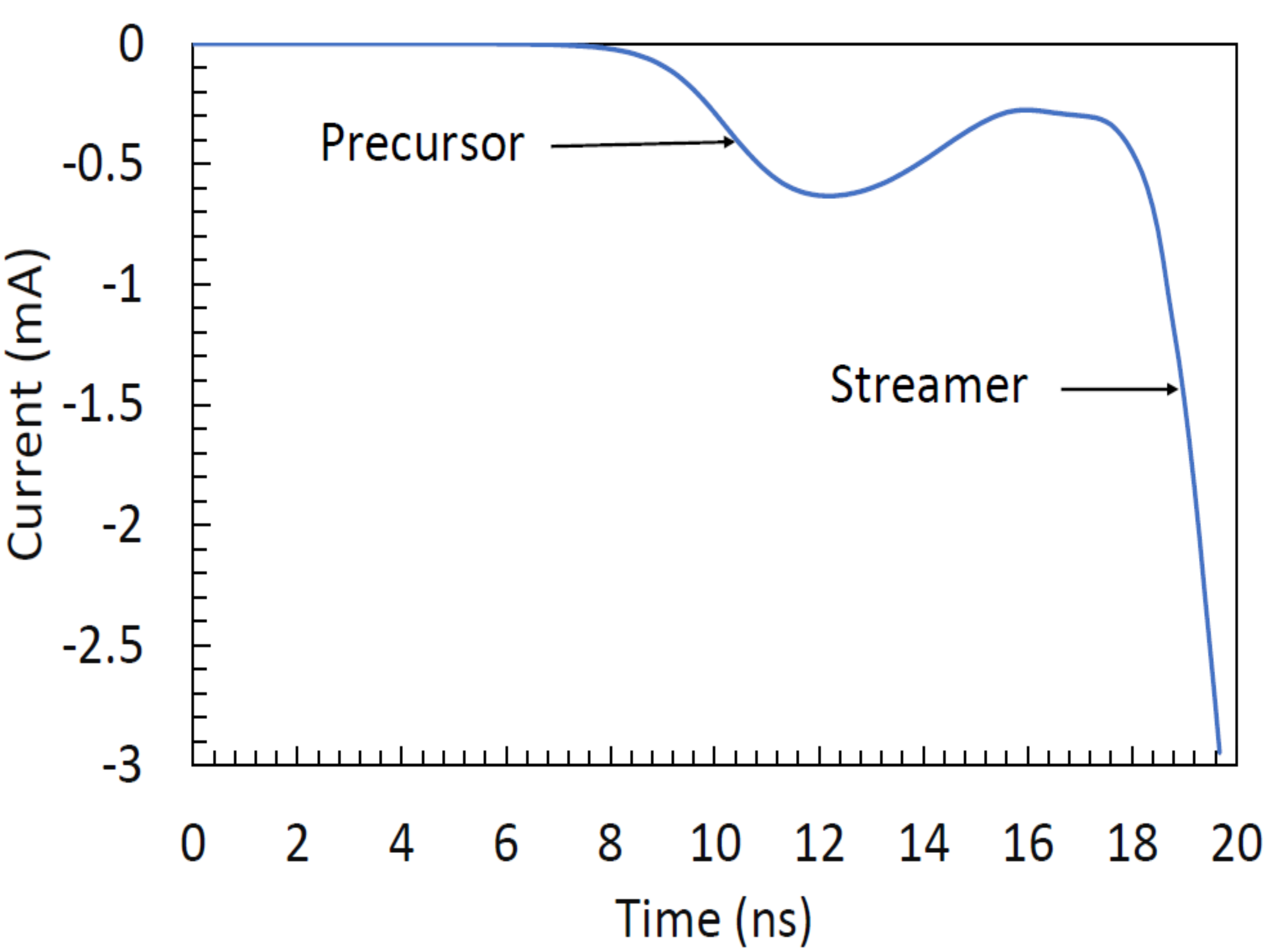}
	\caption{Induced current in case of a streamer event at 47 kV/cm}
	\label{fig:precur}
\end{figure}

The streamers that develop in the operating regime of the RPC are positive when the plasma moves back to the cathode, leading to a rise in the electron density at the cathode. On the other hand, all the electrons leave the gas gap in case of avalanche mode operation. These two conditions have been utilized to identify the streamer and avalanche and subsequently stop the simulation. Figure \ref{fig:aval_grow} and \ref{fig:str_grow} are typical examples of avalanche and streamer events respectively, where the time evolution of the electron density in the gas gap for both the conditions have been depicted. The color code represents the natural logarithm of the electron density in the gas gap. The boundary for each time slice denotes the boundary of the model geometry. In figure \ref{fig:str_grow}, one can see the movement of the higher electron density region towards the cathode as observed experimentally in case of positive streamers.  The aforementioned conditions have been implemented in the following way:
\begin{enumerate}
	\item When the total number of electrons in the gas gap has become less than 1, the simulation has stopped and the case has been identified as avalanche.
	\item When the density of electrons at the cathode has turned out to be non-zero, the simulation has stopped, identifying the mode of operation as streamer.
\end{enumerate}
In figure \ref{fig:efld}, we have shown time evolution of the electric field and electron density for the streamer condition at applied field 47 kV/cm. The background solid color depicts the natural logarithm of the electron density at the two specific instants, mentioned in the figure and the contours represent the electric field. It is clear from the plot of the second instant (at 19 ns) that at the tip of the streamer, the electric field is double of the applied one. It indicates that the space charge field is of same magnitude of the applied field which is one of the characteristics of the streamer condition.\\
Another important characteristic of the streamer event is a precursor which is a comparatively smaller signal occurring before the actual streamer. It has also been observed in the simulation of the induced current (discussed in the next section \ref{sec:4}) and one such case has been shown in figure \ref{fig:precur}.
\section{\label{sec:4} Results}
\begin{figure}[h]
	\centering
	\begin{minipage}{0.48\linewidth}
		\centering
		\includegraphics[width=1.1\linewidth]{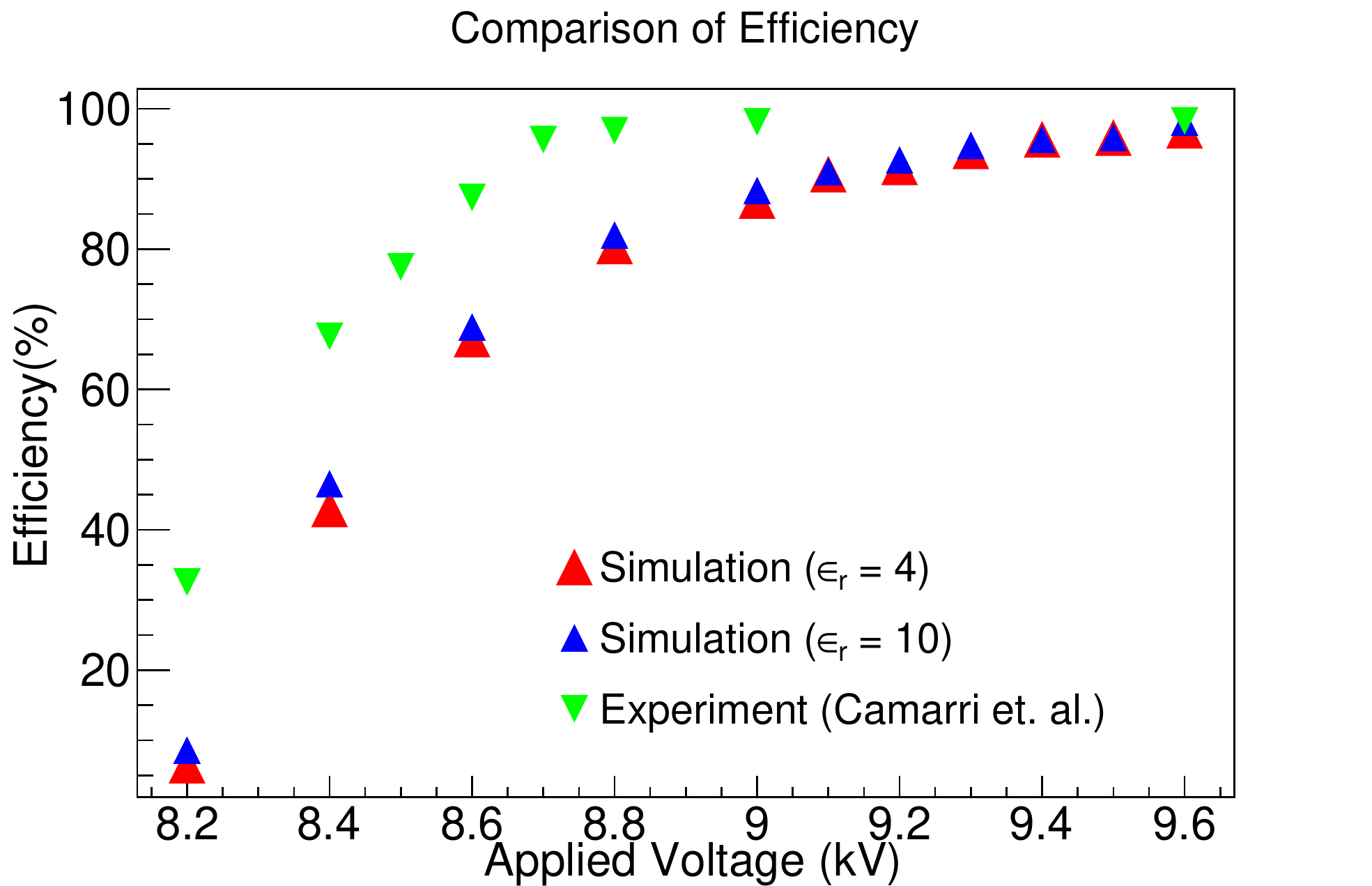}
		\caption{Efficiency as a function of applied voltage}
		\label{fig:efficiency}
	\end{minipage}
	\begin{minipage}{0.5\linewidth}
		\centering
		\includegraphics[width=1.1\linewidth]{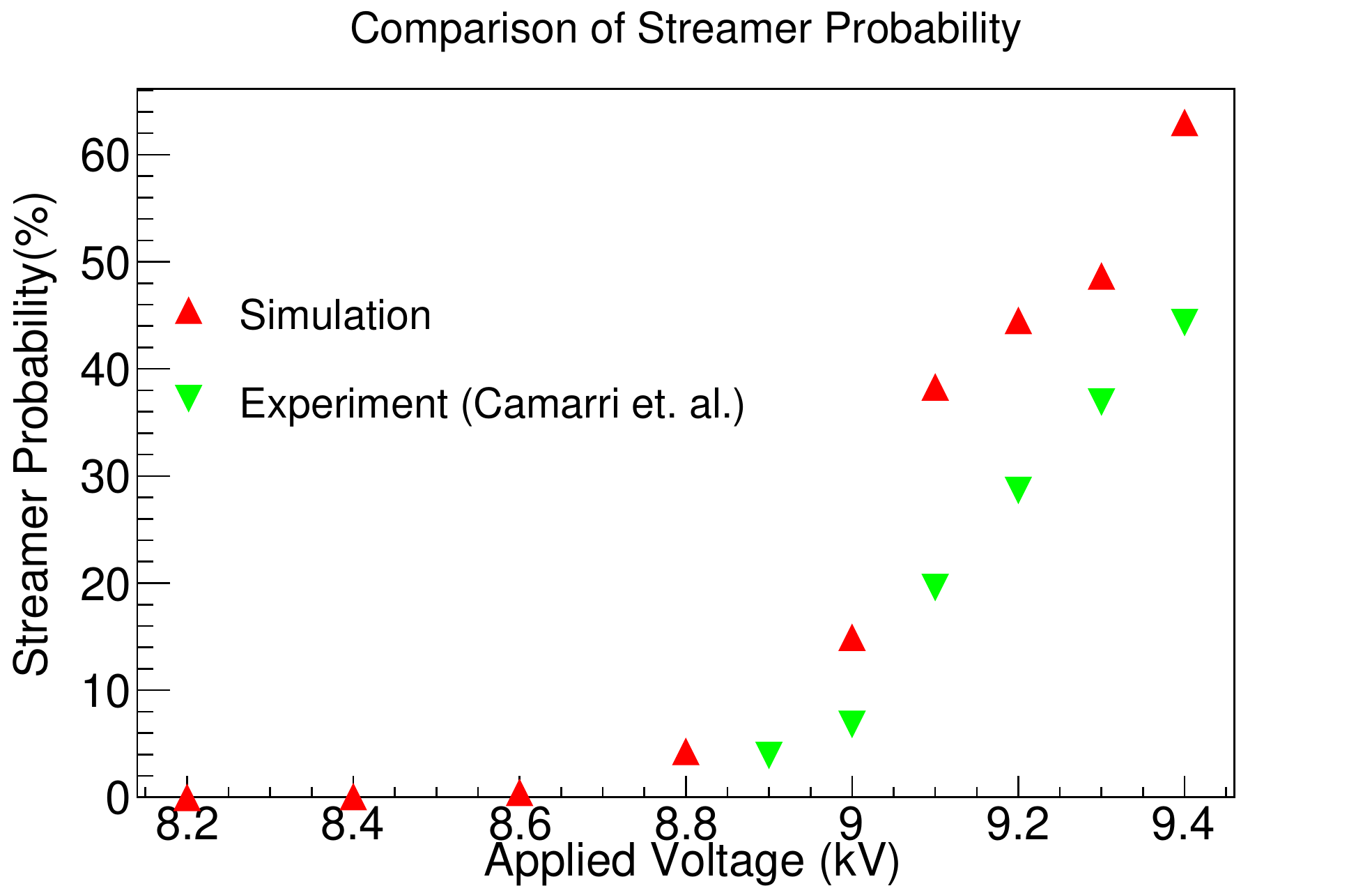}
		\caption{Streamer probability as a function of applied voltage}
		\label{fig:streamer}
	\end{minipage}
\end{figure}

\begin{figure}[h]
		\centering
		\includegraphics[width=0.6\linewidth]{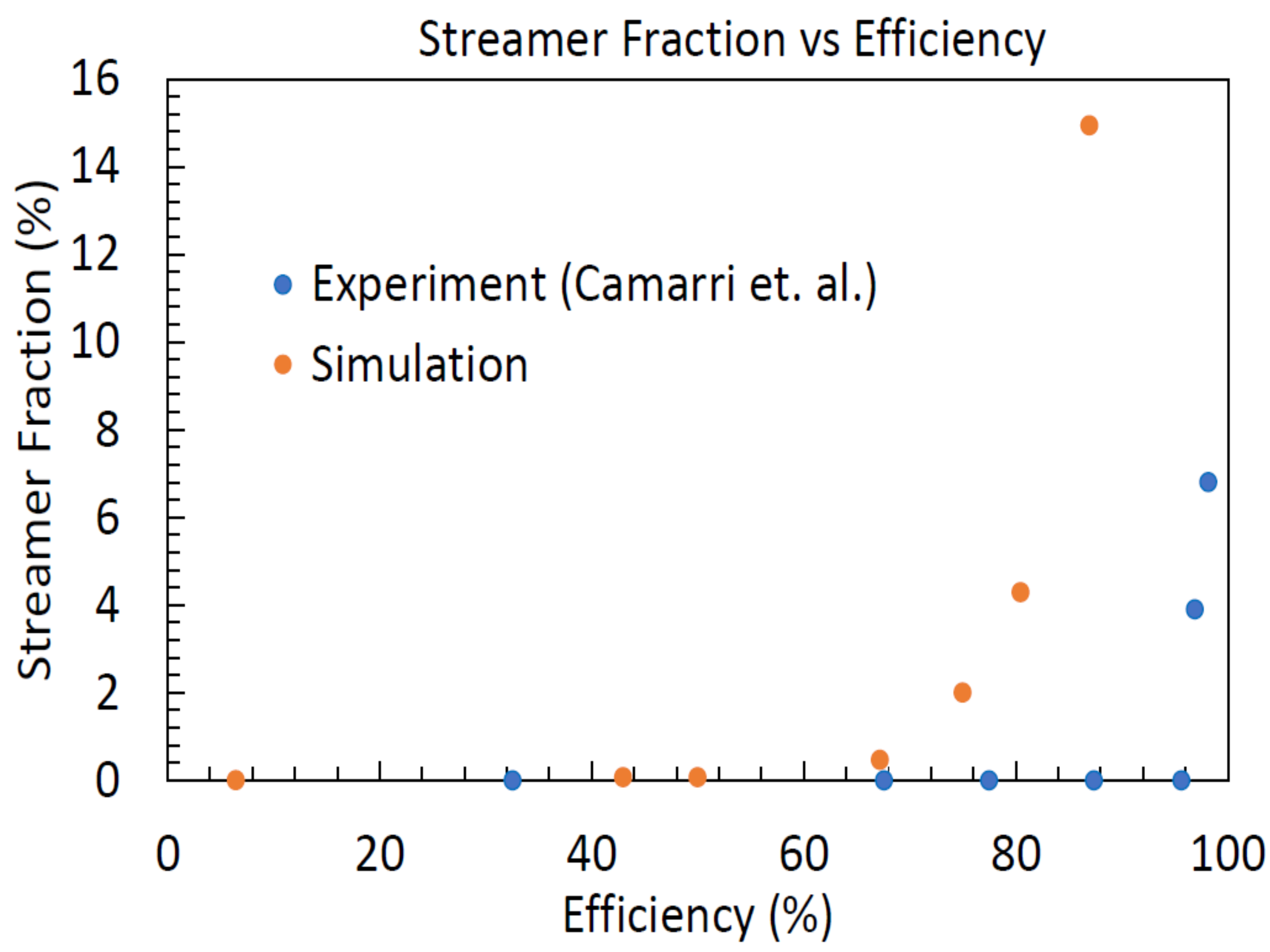}
		\caption{Streamer probability as function of efficiency }
		\label{fig:strfac_eff}
	\end{figure}

For each combination of number and position of seed cluster, the evolution of the charged fluid has been simulated for different applied voltages. Criteria discussed in section \ref{sec:3b} have been used to determine the nature of the RPC signal, as either of avalanche or streamer. In order to compare with the experimental measurements of efficiency and streamer probability reported in \cite{Camarri1998}, the same observables have been estimated using the present simulation model.\\
In our simulation, we have calculated the induced current for each event. The induced current, {\it i(t)}, at an instant, {\it t}, depends upon the $\vec{E}_w$ (weighting field), the electric field in the gas gap when the pick-up electrode of interest is raised to potential $V_w$ while all other electrodes are grounded, $\vec{u}_e(t)$, the instantaneous electron drift velocity, and {\it N(t)}, the number of electrons present at time {\it t} multiplied by the electronic charge, $e_0$.
\begin{eqnarray}\label{Ramo}
i(t)=\frac{\vec{E}_w\vec{u}_e(t)}{V_w}e_0N(t)\\
\frac{E_w}{V_w}=\frac{\epsilon_r}{2b+d\epsilon_r}\label{Ramo2}
\end{eqnarray} The equation \ref{Ramo2} presents the magnitude of the weighting electric field for unit weighting potential which is dependent on the relative permittivity of the electrodes, $\epsilon_r$, and the thicknesses of the electrode and gas gap, b and d, respectively. A threshold criterion following a similar condition mentioned in \cite{Camarri1998} has been used in the present calculation to select the valid signals. In \cite{Camarri1998}, it has been considered that the amplitude of the signal should be greater than 30 mV which is equivalent to 0.1 mV signal acquired across a 25~$\Omega$ resistor. In our work, we have calculated the current corresponding to this threshold criteria and considered it as the threshold to identify the valid events. The size and the mean Z-position of the seed cluster that has given rise to a valid current profile, have been noted as the conditions for selecting the true events. These have been applied to figure \ref{fig:epos} to calculate the total number of valid events. Dividing this number with the total number of events, we have obtained the efficiency for each electric field. In figure \ref{fig:efficiency}, we have compared the simulation and experimental data for the efficiency. The simulation has been carried out for two values of $\epsilon_r$ as that of the resistive material of electrode can vary over the given range.\\
Using the criteria mentioned in the section \ref{sec:3b}, simulation of the seed charge cluster for a group of events at a given applied voltage has been followed to lead to either of the avalanche or streamer. From the simulation, it has been possible to figure out the requisite conditions in terms of the mean Z-position and number of primary electrons for either of them to happen. For example, it has been found that for applied electric field 47 kV/cm, if the seed has 10 electrons, the minimum mean Z-position required for streamer to occur is 1 mm. Therefore, all the seeds with higher value of mean Z-position (away from the anode) and number of primary electrons will lead to streamer discharge at the same electric field. Thus the minimum mean Z-position for each case of seed charge cluster size has been estimated for the occurrence of streamer at each of the applied field. Using these pieces of information, the total number of events leading to streamer discharge has been estimated from the map of the seed cluster size versus mean Z-position of figure \ref{fig:epos}. Dividing it by the number of total events, the streamer probability has been determined. The comparison of the calculated number with the measured value at different voltages has been shown in figure \ref{fig:streamer}. We have also compared the streamer probability as a function of efficiency as calculated from simulation to that observed in the experiment which has been shown in figure \ref{fig:strfac_eff}. \\
It can be noted from the figures \ref{fig:efficiency} and \ref{fig:streamer} that the simulation results have followed the experimental trend quite closely, although a quantitative agreement between them has not been achieved. One of the major reasons may be the lack of information on actual experimental conditions which could not be included in the simulation. On the other hand, the simulation model also needs further investigation to improve its efficacy. 
\section{\label{sec:5}Conclusion}
A numerical simulation framework based on a hydrodynamic model has been developed and used to emulate the dynamics and identify the working mode of an RPC at different applied voltages for a given gas mixture. For validation, the present numerical results have been compared to the published experimental data on efficiency and streamer probability measured with a R134a and butane mixture. Nevertheless, the simulation and the experiment have shown a qualitative disagreement, they have followed similar trend. To improve the efficacy of the simulation model, the authors plan to adopt several measures, such as, reduce the error in the simulation by taking smaller steps in mean Z-position of the seed cluster, smaller steps in total number of primary electrons, reduce the relative tolerance and increase the HEED statistics. Possible reasons of difference between the experimental conditions and the numerical parameters will also be investigated. In addition, the authors have plans to test the simulation model with other available data and also carry out an experiment for the same. However, since the overall trends are in agreement and the estimated streamer probability has been found to be of the same order of magnitude of the measurement, it is expected that the same framework may be useful to study the RPC performance for other gas mixtures.

\acknowledgments

The authors acknowledge INO collaboration and Saha Institute of Nuclear Physics for their help and resources.


\end{document}